\documentclass[12pt,epsf]{article}
\usepackage{amsmath,amssymb}
\usepackage{graphicx}
\usepackage{cite,fancyhdr}
\usepackage{braket}
\usepackage{url}
\usepackage{cancel}
\usepackage{color}
\usepackage{tcolorbox}
\definecolor{light-gray}{gray}{1}
\usepackage{mathtools} 
\usepackage[
colorlinks=true, 
linkcolor=RoyalBlue, 
citecolor=RoyalBlue,
urlcolor=black]{hyperref}
\usepackage[left=3cm,right=3cm,top=3.5cm,bottom=3.5cm,a4paper]{geometry}

\definecolor{Orange}{cmyk}{0,0.61,0.87,0}
\definecolor{JungleGreen}{cmyk}{0.99,0,0.52,0}
\definecolor{OliveGreen}{cmyk}{0.64,0,0.95,0.40}
\definecolor{Brown}{cmyk}{0,0.81,1,0.60}
\definecolor{RoyalBlue}{cmyk}{0.71,0.53,0,0.12}
\definecolor{Gray}{cmyk}{0,0,0,0.40}
\definecolor{Pink}{cmyk}{0.0,1,0,0}
\definecolor{LightPink}{cmyk}{0.0,0.25,0,0}
\definecolor{LLightPink}{cmyk}{0.0,0.10,0,0}
\definecolor{LightBlue}{cmyk}{0.25,0,0,0}
\definecolor{LightGray}{cmyk}{0,0,0,0.2}
\definecolor{LightGreen}{cmyk}{0.3,0,0.3,0}
\def\l{\left}
\def\r{\right}

\def\refheft{\cite{
Feruglio:1992wf,Burgess:1999ha,Giudice:2007fh,Grinstein:2007iv,Alonso:2012px,Buchalla:2012qq,Azatov:2012bz,Contino:2013kra,Jenkins:2013fya,Buchalla:2013rka,Alonso:2014rga,Guo:2015isa,Buchalla:2015qju,Alonso:2017tdy,Buchalla:2017jlu,Buchalla:2018yce}}
\def\refsmeft{\cite{Buchmuller:1985jz, Grzadkowski:2010es, Brivio:2017vri, Manohar:2018aog}}
\begin{document}
\setcounter{footnote}{0}
\setcounter{figure}{0}
\setcounter{table}{0}
\renewcommand{\thefootnote}{\fnsymbol{footnote}}
\numberwithin{equation}{section}
\begin{titlepage}
\begin{flushright}
{\tt 
OU-HET-1134
}
\end{flushright}
\vskip 2.5cm
\begin{center}
{
\large{\bf
Electroweak phase transition \\
in the nearly aligned Higgs effective field theory
}
}
\vskip 1.2cm
Shinya Kanemura{\footnote{
E-mail address:~{\tt{kanemu@het.phys.sci.osaka-u.ac.jp}}}}
,
Ryo Nagai{\footnote{
E-mail address:~{\tt{nagai@het.phys.sci.osaka-u.ac.jp}} or 
{\tt{(r.nagai07@gmail.com)}}}}
and 
Masanori Tanaka{\footnote{
E-mail address:~{\tt{m-tanaka@het.phys.sci.osaka-u.ac.jp}}}}
\vskip 0.4cm
{\small \it
Department of Physics, Osaka University, Toyonaka, Osaka 560-0043, Japan
}\\ [3pt] 
(\today)
\vskip 1.5cm
\begin{abstract}
We investigate the strongly first-order electroweak phase transition
using an effective field theoretical approach. 
The standard effective field theory with finite number truncation of higher dimensional operators fails in the typical parameter space where the strongly first-order phase transition is realized because it cannot describe the non-decoupling quantum effect of new physics beyond the standard model. 
To parameterize the non-decoupling quantum effect, 
we employ the nearly aligned Higgs effective theory 
in which the Higgs potential is parameterized by a Coleman-Weinberg like form. 
Extending this framework with finite temperature corrections,
we study the parameter space for realizing the strongly first-order phase transition, 
and estimate the gravitational wave produced at the phase transition.

\end{abstract}
\end{center}
\end{titlepage}
\renewcommand{\thefootnote}{\roman{footnote}}
\setcounter{footnote}{0}
\section{Introduction}
The electroweak phase transition (EWPT) is one of the most important subjects in particle physics and cosmology. 
This is because the strongly first-order EWPT is required for successful electroweak baryogenesis \cite{Kuzmin:1985mm} 
which explains baryon asymmetry of the Universe.  
Whether the EWPT is strongly first-order or not 
directly depends on 
the structure of the Higgs potential 
whose nature remains unknown yet.   
It is known that the EWPT is crossover in the Standard Model (SM)\cite{Dine:1992vs, Kajantie:1995kf}.
The strongly first-order EWPT requires drastic modification for the Higgs potential of the SM \cite{Kajantie:1996mn, Kajantie:1996qd, Huet:1994jb}.

The strongly first-order EWPT leads to distinctively rich phenomenology. 
One of the most important signatures appears in the Higgs self-coupling. 
The significant deviation in the triple Higgs ($hhh$) coupling from the SM prediction is predicted, which is $\mathcal{O}(10\,\%) \sim \mathcal{O}(100\,\%)$ \cite{Grojean:2004xa,Kanemura:2004ch}.
Therefore, the precision measurement of the $hhh$ coupling at the future collider
experiments is one of the important probes of the strongly first-order EWPT. 
For example, the HL-LHC could reach an accuracy of 50\% \cite{Cepeda:2019klc}, 
while a lepton collider, such as the ILC with high center-of-mass energies, is expected to achieve a precision of some tens of percent \cite{Fujii:2017vwa, Roloff:2019crr}. 
See a complete review, {\it{e.g}}. Ref.~\cite{deBlas:2019rxi}.
In addition to the $hhh$ coupling, 
the strongly first-order EWPT can be tested by observing stochastic gravitational waves from the early universe. 
It has been known that the strongly first-order EWPT predicts the gravitational waves (GWs) in the frequency range of $10^{-3}$\,-\,$10^{-1}\,\mbox{Hz}$ \cite{Grojean:2006bp}, 
which could be measured at the future GW interferometers like LISA \cite{LISA:2017pwj} and DECIGO \cite{Seto:2001qf}.
Furthermore, it has been pointed out that some types of the first-order phase transition could produce primordial black holes (PBHs) \cite{
Kodama:1982sf,
Hawking:1982ga,
Baker:2021nyl,
Kawana:2021tde,
Liu:2021svg,
Jung:2021mku,
Baker:2021sno
}. 
The prediction of the PBHs produced by the strongly first-order EWPT has been recently studied in Ref.~\cite{Hashino:2021qoq,Huang:2022him}.

What kind of the new physics scenario can realize the strongly first-order EWPT? 
The most attractive scenario is the model in which 
the {\it{non-decoupling}} effect appears through the quantum effects of the new physics beyond the SM \cite{
Kanemura:2002vm,
Kanemura:2004mg,
Asakawa:2008se,
Asakawa:2010xj,
Aoki:2012jj,
Kanemura:2016lkz,
Braathen:2019pxr,
Braathen:2019zoh}.
The non-decoupling effect is the heavy new particles' contributions to the low-energy observables 
which are not suppressed by their heavy masses, rather enhanced by power-like contributions of their masses.
The non-decoupling effect realizes the significant modification of the Higgs potential, and leads to the strongly first-order EWPT.
The remarkable feature of this scenario is that we do not need the sizable mixing effects between the new scalars and the SM Higgs boson, which have been severely constrained from the current experiments.
Even if we assume the new particles do not mix with the SM particles,
the non-decoupling effects can appear through the quantum effects if the new particles obtain their masses predominantly from the SM Higgs vacuum expectation value. 
The non-decoupling quantum effect of new physics is therefore essential in the realistic scenarios of the strongly first-order EWPT.
We will be concerned with the EWPT in the non-decoupling new physics scenarios in this paper.
The other scenarios have been studied in Refs.~\cite{Fuyuto:2014yia,Aoki:2021oez} for example.

The importance of the non-decoupling effect for realizing the strongly first-order EWPT has been clarified in many concrete extended Higgs models. 
For example, the non-decoupling effects in the two-Higgs doublet model have been studied in detail and it has shown that the non-decoupling quantum effects from new scalar particles play an important role in realizing the strongly first-order EWPT \cite{
Kanemura:2004ch,
Enomoto:2021dkl, 
Kanemura:2022ozv}.
The detailed analysis of the other extended Higgs models has been also performed and 
it has been shown that 
the non-decoupling quantum effects trigger the strongly first-order EWPT in many new physics scenarios. See Refs.~\cite{
Kakizaki:2015wua, Hashino:2016xoj,Schicho:2021gca,Niemi:2021qvp} for the analysis of the singlet scalar extension of the SM, for example.

Recently the new effective field theory (EFT) describing the non-decoupling quantum effects has been proposed \cite{Kanemura:2021fvp}. (See also Refs.~\cite{
Spannowsky:2016ile,
Reichert:2017puo,
Sondenheimer:2017jin,
Falkowski:2019tft,
Agrawal:2019bpm,
Eichhorn:2020upj,
Cohen:2020xca}.)
The EFT framework is formulated as the modification of the Higgs effective field theory \refheft\, in which the Higgs potential is formulated by a Coleman-Weinberg like form \cite{Coleman:1973jx}.
This EFT systematically describes the new physics in which the non-decoupling effects appear through the radiative correction of new particles whose mixing with the SM particles are much suppressed. 
We call the EFT framework as the ``nearly aligned Higgs effective field theory (naHEFT)''.
The naHEFT should be one of the most simple and systematic formalisms for investigating the strongly first-order EWPT because it can describe the essential part of the non-decoupling quantum effects to the Higgs potential.

In this paper, 
we extend the naHEFT framework with finite temperature corrections, and study the EWPT. 
Our EFT formalism allows us to study the parameter region in which the non-decoupling quantum effects appear, which cannot be treated by the conventional EFT approach consisted of finite number higher dimensional operators. 
Using our EFT, we clarify the parameter region in which the strongly first-order EWPT is realized, and estimate the predicted Higgs self coupling deviations and the GW produced at the phase transition in a model-independent manner.

The rest of this paper is organized as follows: 
In section \ref{sec:formalism}, we explain our formalism. 
We give a brief review of our EFT in section \ref{sec:formalism-T0}, and then introduce the prescription for introducing the finite temperature corrections in section \ref{sec:formalism-finiteT}.
We study the EWPT in section \ref{sec:strongly first-order EWPT}.
In section \ref{sec:vsSMEFT}, we discuss the validity of the conventional EFT approach in the analysis of the strongly first-order EWPT.
In section \ref{sec:GWs}, we estimate the GW spectra produced at the strongly first-order EWPT. 
We discuss the phenomenology of the strongly first-order EWPT in section \ref{sec:result}.
Finally we summarize our results and conclusions in section \ref{sec:conclusion}.

\section{Our formalism}
\label{sec:formalism}
\subsection{EFT for non-decoupling new physics}
\label{sec:formalism-T0}
We consider the physics beyond the SM in which all new particles are heavier than the SM particles.
In this case, we can discuss the new physics effects to the low-energy observables in a model-independent way by using an effective field theoretical approach. 
We focus on the Higgs potential in the effective field theory,
\begin{align}
V_{\rm{EFT}}(\Phi)
\,=\,
V_{\rm{SM}}(\Phi)
\,+\,
V_{\rm{BSM}}(\Phi)\,,
\end{align}
where the new particles are integrated out. 
We decompose the effective Higgs potential into the SM part ($V_{\rm{SM}}(\Phi)$)
and the new physics part ($V_{\rm{BSM}}(\Phi)$).
Here $\Phi$ denotes the SM Higgs doublet field which contains the physical $125$\,GeV Higgs boson field $h$ as $|\Phi|^2=(v+h)^2/2$ with $v\simeq 246\,\mbox{GeV}$.
Ignoring the radiative corrections, the SM part can be parameterized as 
\begin{align}
V_{\rm{SM}}(\Phi)
\,=\,
m^2\, |\Phi|^2
\,+\,
\lambda\, |\Phi|^4
\,,
\end{align}
where both $m^2$ and $\lambda$ are real parameters.

There are several possible forms of the new physics part, on the other hand.
The most general form is the Higgs effective field theory (HEFT) framework \refheft\,
in which the new physics effects to the Higgs potential are parameterized as
\begin{align}
V_{\rm{BSM}}(\Phi)
\,=\,
f(|\Phi|^2)
\qquad (\mbox{HEFT})
\,,
\label{eq:VHEFT}
\end{align}
with $f(|\Phi|^2)$ being an arbitral function of $|\Phi|^2$. 
The remarkable feature of the HEFT formalism is that the function $f(|\Phi|^2)$ can contain operator which is not only analytic but also non-analytic at $\Phi=0$, such as $\sqrt{|\Phi|^2}$, $\log|\Phi|^2$ and so on. 
As clarified in Refs.~\cite{
Alonso:2016oah, 
Falkowski:2019tft,
Cohen:2020xca,
Cohen:2021ucp,
Kanemura:2021fvp}, 
the non-analytic Higgs operators parameterize the non-decoupling new physics effects to the Higgs sector.

Another widely used EFT framework is so-called the standard model effective field theory (SMEFT) \refsmeft,
in which $f(|\Phi|^2)$ is restricted to be a polynomial in $|\Phi|^2$ as
\begin{align}
V_{\rm{BSM}}(\Phi)
\,=\,
\sum_{n=1} \,c_n\, |\Phi|^{2n}
\qquad (\mbox{SMEFT})
\,,
\label{eq:VSMEFT}
\end{align}
with $c_n$ being constant parameters.
The SMEFT can be applied if all new physics effects are categorized as the decoupling effects. 
We will clarify the validity of the SMEFT framework later.

Which EFT should we employ to investigate the strongly first-order EWPT?
The important knowledge to answer this question is that 
the non-decoupling new physics effects play important roles in realizing the strongly first-order EWPT in many extended Higgs models.
For example, in the two Higgs doublet model, the strongly first-order EWPT is typically triggered by the non-decoupling quantum effects from the heavy extra scalars \cite{Kanemura:2004ch,Enomoto:2021dkl, Kanemura:2022ozv}. The same physics happens in the other extended Higgs models such as $O(N)$ singlet extended model \cite{Kakizaki:2015wua,Hashino:2016xoj} and so on.
These observations lead us to use the EFT framework which can systematically describe the non-decoupling quantum effects. 
The HEFT framework (\ref{eq:VHEFT}) is one of the candidates because it can express the non-decoupling effects by non-analytic Higgs operators.  However, it is not straightforward to apply the HEFT to the phenomenological analysis because the parameter space of the HEFT spans very broadly due to its generality. 
We therefore introduce a new EFT framework in which 
non-decoupling new physics effects are more systematically described.
The Higgs potential in this framework is parameterized as
\begin{align}
V_{\rm{BSM}}(\Phi)
\,=\,
\frac{\xi}{4}\,\kappa_0\,
[\mathcal{M}^2(\Phi)]^2
\ln\frac{\mathcal{M}^2(\Phi)}{\mu^2}
\,,
\label{eq:VnaHEFT}
\end{align}
where $\kappa_0$ and $\mu^2$ are real dimension-less and mass dimension two parameters, respectively. $\xi=1/(4\pi)^2$. $\mathcal{M}^2(\Phi)$ is an arbitral function of $|\Phi|^2$, whose mass dimension is two.
Our parameterization is inspired by the observation from the concrete new physics model studies where the non-decoupling new physics effect to the Higgs potential is described by a Coleman-Weinberg like form \cite{Coleman:1973jx}. 
In Ref.~\cite{Kanemura:2021fvp}, 
we have formulated the EFT description for not only the Higgs potential but also the other parts, and we call our EFT formalism as ``nearly aligned Higgs effective field theory (naHEFT)''.  

In this paper, we employ the naHEFT framework (\ref{eq:VnaHEFT}) to investigate the strongly first-order EWPT.
For simplicity, we assume $\mathcal{M}^2(\Phi)$ is parameterized as
\begin{align}
\mathcal{M}^2(\Phi)
\,=\,
M^2 + \kappa_{\rm{p}}\, |\Phi|^2
\,,
\label{eq:Msq}
\end{align}
where $M^2$ and $\kappa_{\rm{p}}$ are real parameters.
It is straightforward to generalize the structure of $\mathcal{M}^2(\Phi)$. 
Note that our EFT (\ref{eq:VnaHEFT}) can describe the non-decoupling effect thanks to the presence of $\ln (\mathcal{M}^2(\Phi)/\mu^2)$ term.
For later use, we introduce $\Lambda$ and $r$ as
\begin{align}
\Lambda^2 
&\,=\,
M^2 + \frac{\kappa_{\rm{p}}}{2}v^2
\,,\\
r
&\,=\,
\frac{\frac{\kappa_{\rm{p}}}{2}v^2}{\Lambda^2}
=
1-\frac{M^2}{\Lambda^2}
\,.
\end{align}
$\Lambda$ should be interpreted as the physical mass of the integrated new particle.
The scale $\Lambda$ can be therefore regarded as the cutoff scale of our EFT.
In what follows, we assume $\Lambda > v$ to work the EFT description. 
The parameter $r$ parameterizes ``non-decouplingness'' of the integrated new particles.
A non-decoupling case corresponds to the case with $|r|\simeq \mathcal{O}(1)$.
The scenarios with classically scale invariance
\cite{
Gildener:1976ih,
Takenaga:1993ux,
Funakubo:1993jg,
Bardeen:1995kv,
Lee:2012jn,
Ishiwata:2011aa,
Guo:2014bha,
Farzinnia:2014yqa,
Endo:2015ifa,
Fuyuto:2015jha,
Endo:2015nba,
Hashino:2015nxa,
Haba:2015lka,
Okada:2015gia,
Helmboldt:2016mpi,
Hashino:2016rvx,
Hashino:2016xoj,
Fujitani:2017gma,
Prokopec:2018tnq,
Jung:2019dog,
Braathen:2020vwo}, in which the sizable non-decoupling effects appear, 
can be described by taking $r=1$ (corresponding to $M^2=0$) and $m^2=0$ in our EFT framework.

In our analysis, we impose
\begin{align}
&\frac{dV_{\rm{EFT}}}{dh}\biggl|_{h=0}
\,=\,
0\,,
\label{eq:renom-cond1}
\\
&
\frac{d^2V_{\rm{EFT}}}{dh^2}\biggl|_{h=0}
\,=\,
M^2_h\,,
\label{eq:renom-cond2}
\end{align}
where $M_h$ denotes the SM Higgs mass, $M_h\simeq 125\,\rm{GeV}$. Using the conditions (\ref{eq:renom-cond1}) and (\ref{eq:renom-cond2}), we can express $V_{\rm{EFT}}$ in terms of $r$, $\Lambda$, $\kappa_0$, $v$, and the masses of the SM particles.  

In our EFT, the Higgs self-couplings are modified from the SM prediction. For example, the triple Higgs coupling
\begin{align}
V_{\rm{EFT}}
\,\ni \, 
\frac{1}{3!}\frac{3M^2_h}{v}
\,\kappa_3\,
h^3
\,,
\end{align}
is given as \cite{Kanemura:2021fvp}
\begin{align}
\kappa_3
\,=\,
1+
\frac{4}{3}\,\xi\,
\frac{\Lambda^4}{v^2 M^2_h}\, \kappa_0 \,r^3
\,.
\label{eq:kappa3}
\end{align}
$\kappa_3=1$ at the tree-level in the SM. 
We here ignore the corrections due to the wave function renormalization of the Higgs field,
which does not modify the leading contribution to $\kappa_3$. 
See Ref.~\cite{Kanemura:2021fvp} for the complete expression.
We note that, when $|r|\simeq 1$, the deviation of the triple Higgs coupling from SM prediction ($=\kappa_3-1$) is enhanced by the forth power of the new physics scale, $\Lambda$. This is nothing but the non-decoupling new physics effect to the Higgs self-coupling. 
The sizable modifications of the Higgs self-couplings typically correlate with the strongly first-order EWPT as we will show later.

When $r\simeq 0$ (corresponding to $M^2\simeq \Lambda^2$), the new physics effect should be decoupled because the new particle obtains the mass almost independently from the Higgs vacuum expectation value. 
In this case, our EFT falls into the SMEFT form (\ref{eq:VSMEFT}).
Let us check this explicitly. 
We first note that, when $M^2\neq 0$ (corresponding to $r\neq 1$), $\ln \mathcal{M}^2(\Phi)/\mu^2$ in Eq.~(\ref{eq:VnaHEFT}) can be decomposed as
\begin{align}
\ln \frac{\mathcal{M}^2(\Phi)}{\mu^2}
\,=\,
\ln\frac{M^2}{\mu^2}
+
\ln\l(1+x_\Phi\r),
\end{align}
where 
\begin{align}
x_\Phi
\,=\,
\frac{r}{1-r}\frac{|\Phi|^2}{\frac{v^2}{2}}
\,.
\end{align}
Note that $x_\Phi \ll 1$ when $r\simeq 0$ and $|\Phi|\lesssim v$. 
If $x_\Phi \ll 1$, we can expand $\ln(1+x_\Phi)$ as
\begin{align}
\ln\l(1+x_\Phi\r)
\,=\,
x_\Phi
-
\frac{x^2_\Phi}{2}
+
\frac{x^3_\Phi}{3}
+
\mathcal{O}(x^4_\Phi)
\,.
\label{eq:log-expand}
\end{align}
Substituting the right-handed side of Eq.~(\ref{eq:log-expand}) into the original effective potential (\ref{eq:VnaHEFT}),
we obtain the effective potential with the SMEFT form (\ref{eq:VSMEFT}).

If we truncate $\ln(1+x_\Phi)$ at $\mathcal{O}(x_\Phi)$, 
our effective potential (\ref{eq:VnaHEFT}) can be expressed by the SMEFT form (\ref{eq:VSMEFT}) up to mass dimension six operators.
Imposing the conditions Eqs.~(\ref{eq:renom-cond1}) and (\ref{eq:renom-cond2}), 
the new physics contribution is obtained as
\begin{align}
V_{\rm{BSM}}(\Phi)
\,= \,
\frac{1}{f^2}\l(|\Phi|^2-\frac{v^2}{2}\r)^3
\,,
\label{eq:VnaHEFT-dim6}
\end{align}
where $f$ is given as
\begin{align}
\frac{1}{f^2}\,=\,
\frac{2}{3}\,\xi\,\kappa_0\, \frac{\Lambda^4}{v^6}\frac{r^3}{1-r}
\,.
\end{align}
The decoupling limit corresponds to $r\to 0$, which leads to $f\to \infty$.

On the other hand, if we truncate $\ln(1+x_\Phi)$ at $\mathcal{O}(x^2_\Phi)$, 
our effective potential is expressed as the SMEFT form (\ref{eq:VSMEFT}) up to mass dimension eight operators. Using Eqs.~(\ref{eq:renom-cond1}) and (\ref{eq:renom-cond2}) again, we find 
\begin{align}
V_{\rm{BSM}}(\Phi)
\,= \,
\frac{1}{f^2_6}\l(|\Phi|^2-\frac{v^2}{2}\r)^3
-
\frac{1}{f^4_8}\l(|\Phi|^2-\frac{v^2}{2}\r)^4
\,,
\label{eq:VnaHEFT-dim6+8}
\end{align}
where $f_6$ and $f_8$ are given as
\begin{align}
&\frac{1}{f^2_6}\,=
\frac{1}{f^2}\,
\frac{1-2r}{1-r}
\,,\\
&\frac{1}{f^4_8}\,=\,
\frac{1}{2 f^2 v^2} \frac{r}{1-r}
\,.
\end{align}
The decoupling limit again corresponds to $r\to 0$, which leads to $f_{6,8}\to \infty$.

In the same way, we obtain the SMEFT form up to mass dimension ten operators by truncating $\ln(1+x_\Phi)$ at $\mathcal{O}(x^3_\Phi)$. We find
\begin{align}
V_{\rm{BSM}}(\Phi)
\,= \,
\frac{1}{F^2_6}\l(|\Phi|^2-\frac{v^2}{2}\r)^3
-
\frac{1}{F^4_8}\l(|\Phi|^2-\frac{v^2}{2}\r)^4
+
\frac{1}{F^6_{10}}\l(|\Phi|^2-\frac{v^2}{2}\r)^5
\,,
\label{eq:VnaHEFT-dim6+8+10}
\end{align}
where $F_6$, $F_8$, and $F_{10}$ are given as
\begin{align}
&\frac{1}{F^2_6}
\,=\,
\frac{1}{f^2} \frac{3r^2-3r+1 }{(1-r)^2}
\,,\\
&\frac{1}{F^4_8}
\,=\,
\frac{1}{2 f^2 v^2}\frac{r(1-3r)}{(1-r)^2}
\,,\\
&\frac{1}{F^6_{10}}
\,=\,
\frac{2}{5 f^2 v^4} \left( \frac{r}{1-r} \right)^2
\,.
\end{align}
We note that $F_{6,8,10}\to \infty$ when $r\to 0$.

We emphasize that the SMEFT approximations (\ref{eq:VnaHEFT-dim6}), (\ref{eq:VnaHEFT-dim6+8}), and (\ref{eq:VnaHEFT-dim6+8+10}) fail when $|x_\Phi|\simeq1$, which typically corresponds to the non-decoupling case. We will revisit this observation later.

\subsection{Finite temperature corrections from new physics}
\label{sec:formalism-finiteT}
Let us next discuss how to turn on finite temperature corrections 
in our EFT formalism.
We first note that $V_{\rm{BSM}}$ (\ref{eq:VnaHEFT}) can be expressed as 
\begin{align}
V_{\rm{BSM}}(\Phi)
\,=\,
\frac{\kappa_0}{2}
\int\frac{d^4 k_E}{(2\pi)^4}
\ln\biggl[
k^2_E+\mathcal{M}^2(\Phi)
\biggr]
\,,
\end{align} 
where $k_E$ denoted a Euclidian momentum.
We here neglect the ultraviolet divergent and the constant terms because they are irrelevant to the following argument.
We can turn on the finite temperature corrections from the new physics by performing the following replacement \cite{PhysRevD.9.3312,PhysRevD.9.3320,PhysRevD.9.3357},
\begin{align}
{{
\int\frac{d^4 k_E}{(2\pi)^4}
\ln
\biggl[
k^2_E+\mathcal{M}^2(\Phi)
\biggr]}}
~\to~
T
\sum_{n=-\infty}^{\infty}
\int\frac{d^{3} k_E}{(2\pi)^{3}}
\ln
\biggl[
\vec{k}^2_E+w^2_{n}+\mathcal{M}^2(\Phi)
\biggr]
\,,
\end{align}
where $T$ is temperature, and $w_{n}$ denotes Matsubara frequency for the integrated new particle,
\begin{align}
w_{n}=
\begin{dcases}
2n\pi\, T
& \kappa_0>0\\
(2n+1)\,\pi T
& \kappa_0<0
\end{dcases}
\,.
\end{align} 
The new physics effect with the finite temperature correction is therefore obtained as
\begin{align}
{{V_{\rm{BSM}}(\Phi,T)}}
\,= \,
\frac{\kappa_0}{2}
\,T
\sum_{n=-\infty}^{\infty}
\int\frac{d^3 k_E}{(2\pi)^3}
\ln\biggl[
\vec{k}^2_E+w^2_n+\mathcal{M}^2(\Phi)
\biggr]
\,.
\label{eq:finitetint}
\end{align} 
We note that the above prescription for introducing the finite temperature correction can be applied for general $\mathcal{M}^2(\Phi)$, not only for the specific form (\ref{eq:Msq}).

We can decompose the right-handed side of Eq.~(\ref{eq:finitetint})
into a zero-temperature part (\ref{eq:VnaHEFT}) and a $T$-dependent part ($\Delta V_{\rm{BSM},T}(\Phi,T)$) as,
\begin{align}
V_{\rm{BSM}}(\Phi,T)
&
\,=\,
V_{\rm{BSM}}(\Phi)
\,+\,
\Delta V_{\rm{BSM},T}(\Phi,T)
\,,
\end{align}
where 
\begin{align}
\Delta V_{\rm{BSM},T}(\Phi,T)
&\,=\,
8\,\xi\, T^4\,
\kappa_0\,
J_{\rm{BSM}}\l(\frac{\mathcal{M}^2(\Phi)}{T^2}\r)
\,.
\label{eq:thermalint}
\end{align}
The function $J_{\rm{BSM}}$ is defined as
\begin{align}
{{J_{\rm{BSM}}\l(\frac{\mathcal{M}^2(\Phi)}{T^2}\r)}}
\,=\,
\int^\infty_0 dk^2
k^2
\biggl(
1-\mbox{sign}(\kappa_0)\,
e^{-\sqrt{k^2+\frac{\mathcal{M}^2(\Phi)}{T^2}}}
\biggr)
\,,
\end{align}
with
\begin{align}
\mbox{sign}(\kappa_0)=
\begin{dcases}
1
& \kappa_0>0\\
-1
& \kappa_0<0
\end{dcases}
\,.
\end{align}

It should be noted that, in the case with $\kappa_0>0$, the finite temperature correction from $n=0$ mode suffers from infrared (IR) divergence. In order to regularize the IR divergence, we need to introduce the screening effects from the thermal corrections appropriately.
In our analysis, we employ the ``Parwani prescription \cite{Parwani:1991gq}'' to regularize the IR divergence. This prescription requires the following replacement,
\begin{align}
\mathcal{M}^2(\Phi)
~\to~
\hat{\mathcal{M}}^2(\Phi,T)
\,=\,
\mathcal{M}^2(\Phi)\,+\,\Pi_{\rm{BSM}}(T)
\,,
\end{align}
with $\Pi_{\rm{BSM}}(T)$ being the Debye mass of the integrated new bosonic particle{\footnote{
For example, a detailed treatment of the thermal corrections is discussed in Ref. \cite{Croon:2020cgk}.
}}. Unfortunately, we cannot determine $\Pi_{\rm{BSM}}(T)$ without specifying the underlying theory. In our analysis, we simply take{\footnote{
We take this form because, in the singlet extended model where the new singlet $S$ interacts with the SM Higgs field $\Phi$ through $\mathcal{L}_{\rm{int}}= \frac{\kappa_{\rm{p}}}{2}S^2|\Phi|^2$, the Debye mass of the singlet is obtained as $\Pi_S = \frac{\kappa_{\rm{p}}}{6}\,T^2\,\Theta(\kappa_0)$. 
}}
\begin{align}
\Pi_{\rm{BSM}}(T)
\,=\,
\frac{\kappa_{\rm{p}}}{6}\,T^2\,
\Theta(\kappa_0)
\,,
\label{eq:PiX}
\end{align}
where
\begin{align}
\Theta(\kappa_0)=
\begin{dcases}
1
& \kappa_0>0\\
0
& \kappa_0\leq 0
\end{dcases}
\,.
\end{align}

We remark that, for the decoupling case ($r\simeq0$), 
the finite temperature correction from the BSM particle (\ref{eq:thermalint}) should be much suppressed. 
This is because, when $r\simeq0$, $\mathcal{M}^2(\Phi)\simeq \Lambda^2$ and the scale $\Lambda$ should be much larger than the typical temperature to work the EFT description. 
The scale hierarchy $\mathcal{M}^2(\Phi)\simeq \Lambda^2 \gg T^2$ eventually induces the Boltzmann suppression as 
\begin{align}
J_{\rm{BSM}}\,\biggl|_{r\simeq 0}
\,\propto\, 
e^{-\sqrt{\frac{\mathcal{M}^2(\Phi)}{T^2}}}
\,\simeq\,
 e^{-\frac{\Lambda}{T}}\ll 1
\,. 
\end{align}
It is therefore hard for the SMEFT formalism to implement the sizable finite temperature corrections in a self-consistent way.

On the other hand, for the non-decoupling case ($|r|\simeq 1$), the finite temperature correction (\ref{eq:thermalint}) can induce the sizable contribution to the phase transition. The most important contribution for realizing the strongly first-order EWPT can be obtained when $\kappa_0>0$ and $r\simeq 1$. In this case, we obtain $\phi^3$ term with $\phi$ being the order parameter in the effective potential as
\begin{align}
\Delta V_{\rm{BSM}}(\Phi,T)\biggl|_{\kappa_0>0,\,r\simeq 1} 
\,\ni\,
 -E_{\rm{BSM}}\,T\,\phi^3\,,
\end{align}
where
\begin{align}
E_{\rm{BSM}}
\,=\,\frac{4\pi}{3}\,\xi\,\kappa_0\, \frac{\Lambda^3}{v^3}
\,.
\end{align}
We will clarify the importance of the non-decoupling effect for realizing the strongly first-order EWPT by the numerical analysis performed below.

\section{Strongly first-order EWPT}
\label{sec:strongly first-order EWPT}
We next investigate the parameter space in which the strongly first-order EWPT is realized in our EFT framework. 
The parameter set in our EFT analysis is
\begin{align}
\kappa_0\,,\qquad
\Lambda\,,\qquad
r\,.
\end{align}
In what follows, we focus on the $\kappa_0>0$ case because we obtain the $\phi^3$ correction in this case as we discussed, which should play an important role in realizing the strongly first-order EWPT.
Moreover, we focus on the parameter space where $0 \leq r \leq 1$.
The case with $r=0$ and $1$ correspond to the case where $M^2=\Lambda^2$ (completely decoupling case) and $M^2=0$ (maximally non-decoupling case), respectively.
We use {\tt{CosmoTransitions}} package \cite{Wainwright:2011kj} to calculate the tunneling rate of the scalar field during the phase transition.
In our analysis, we include the SM one-loop corrections in addition to the new physics correction.

\begin{figure}[t]
	\centering
	\includegraphics[width=7cm,clip]{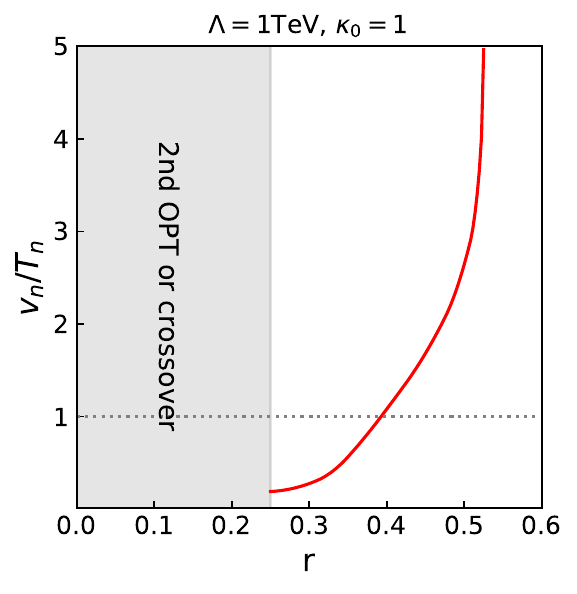}		
	\caption{ 
		The relation between $v_n/T_n$ and the non-decouplingness $r$. 
		We take $\Lambda=1\,\mbox{TeV}$ and $\kappa_0=1$.
		}
	\label{fig:vnTn}
\end{figure}
First, we show the relation between $v_n/T_n$ and the non-decouplingness $r$ in figure \ref{fig:vnTn}. 
Here we take $\Lambda=1\,\mbox{TeV}$ and $\kappa_0=1$.
$v_n$ and $T_n$ are the vacuum expectation value and the temperature when the vacuum bubble is nucleated.
The bubble nucleation temperature $T_n$ is defined as the temperature when $S_3(T_n)/T_n\simeq 140$ with $S_3$ being the three-dimensional Euclidian action. The strongly first-order EWPT is realized when $v_n/T_n\gtrsim 1$.
We find that the strongly first-order EWPT can be realized when $r$ is sizable as we expected.
For example, if $r\gtrsim 0.39$, the strongly first-order EWPT can be realized when $\Lambda=1\,\mbox{TeV}$ and $\kappa_0=1$. 
We note that the large $r$ induces the divergence of $v_n/T_n$.
This is because, if we take too large $r$, the phase transition is not completed.
Therefore, the non-decouplingness $r$ should be bounded by a certain critical value to realize the electroweak symmetry breaking. 
We numerically estimated this upper bound and confirmed that this bound is almost compatible with the vacuum stability constraint at zero-temperature which was derived in Ref.~\cite{Kanemura:2021fvp}.
We also remark that, for extremely small $r$,  we cannot realize the first-order phase transition because the new physics effect becomes decoupled. The gray shaded region in figure \ref{fig:vnTn} corresponds to the region in which the first-order phase transition cannot be realized. 
We find that $r$ is required to be larger than $\simeq 0.25$ to realize the first-order phase transition when $\Lambda=1\,\mbox{TeV}$ and $\kappa_0=1$.

\begin{figure}[t]
	\centering
	\includegraphics[width=7cm,clip]{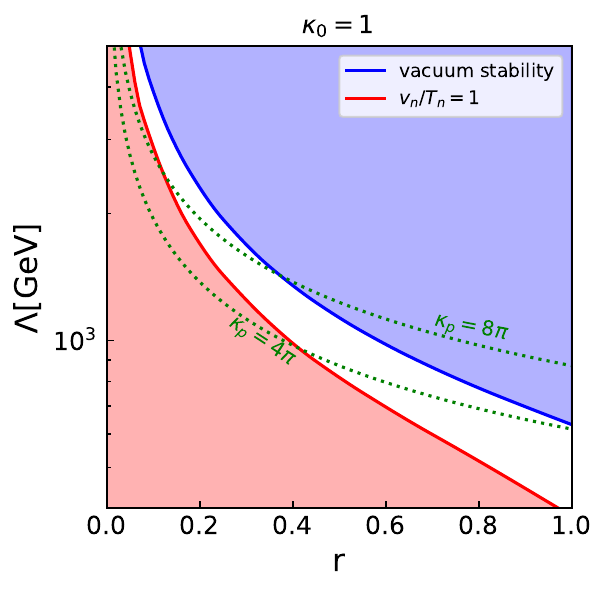}
	~~
	\includegraphics[width=7cm,clip]{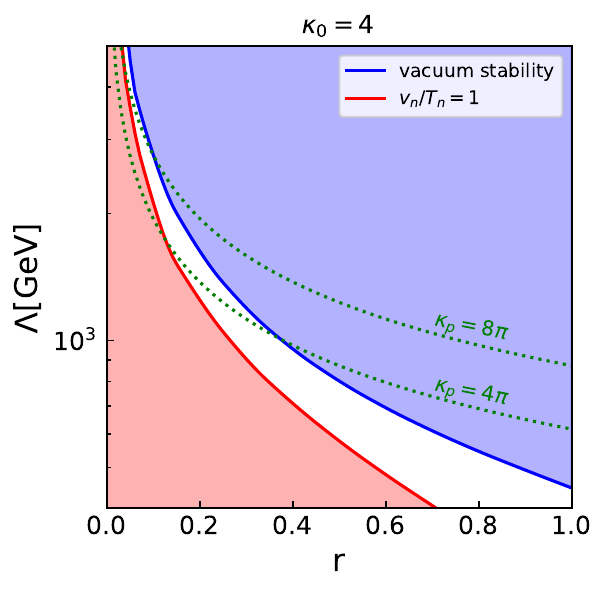}		
	\caption{ 
		The parameter region for the strongly first-order EWPT. 
		The left and right figures correspond to the case with $\kappa_0 = 1$
		and $\kappa_0=4$, respectively.
		The strongly first-order EWPT can be realized in the white region.
		$v_n/T_n < 1$ or the first-order EWPT cannot be realized in the red region.
		The blue region is excluded by the vacuum stability 
		which is estimated in Ref.~\cite{Kanemura:2021fvp}
				}
	\label{fig:SFOEWPT}
\end{figure}
Figure \ref{fig:SFOEWPT} shows the parameter space for the strongly first-order EWPT. 
Here we take $\kappa_0 = 1$ and $4$.
The white region corresponds to the parameter region in which $v_n/T_n>1$. 
In the red region, $v_n/T_n < 1$ or the first-order EWPT cannot be realized.
We find that sizable $r$ expands the parameter region for the strongly first-order EWPT.
We have confirmed that our EFT results agree with the results of the analysis in concrete extended Higgs models \cite{
Kanemura:2004ch,
Kakizaki:2015wua,
Hashino:2016xoj,
Kanemura:2022ozv
}.
The blue region is excluded by the vacuum stability constraint in which the electroweak vacuum is no longer global minimum at $T=0$. We also plot the contours of $\kappa_{\rm{p}}=4\pi$ and $8\pi$. 
Considering the perturbative unitarity constraint on the scattering amplitudes involving the integrated new particles, 
one can obtain $\kappa_{\rm{p}}\lesssim 8\pi/\sqrt{|\kappa_0|}$ \cite{Kakizaki:2015wua} with an $\mathcal{O}(1)$ ambiguity coming from the ambiguity of the prescription of the unitarity constraint.

\section{Validity of the finite number truncation of higher dimensional operators}
\label{sec:vsSMEFT}
Let us next discuss the validity of the description with the finite number truncation of higher dimensional operators (\ref{eq:VSMEFT}). 
As we discussed, the approximation with the finite number truncation of higher dimensional operators does not work when $|x_\Phi|\sim 1$ which typically corresponds to the non-decoupling case.
In this section, we estimate the validity of the finite number truncation in the typical parameter space of the strongly first-order EWPT.

\begin{figure}[t]
	\centering
	\includegraphics[width=7cm]{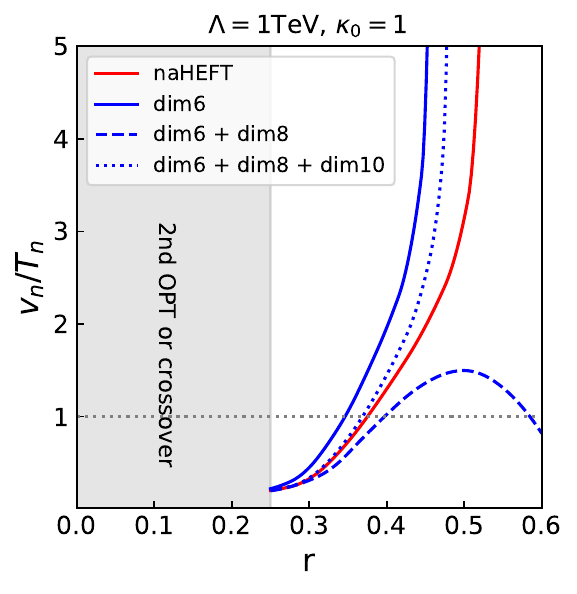}
	\caption{ 
	      $v_n/T_n$ estimated by our EFT without finite temperature corrections from new physics (red), 
	      the effective potential approximated up to the dimension six (blue),
	     eight (dashed blue), and ten (dotted blue) operators. 
	      We take $\Lambda=1\,\mbox{TeV}$ and $\kappa_0=1$.
				}
	\label{fig:vnTn-vsSMEFT}
\end{figure}
In figure \ref{fig:vnTn-vsSMEFT},
we compare our EFT results of $v_n/T_n$ with ones estimated by the effective potential approximated up to mass dimension six (\ref{eq:VnaHEFT-dim6}), mass dimension eight (\ref{eq:VnaHEFT-dim6+8}), and mass dimension ten (\ref{eq:VnaHEFT-dim6+8+10}). 
In order to compare our EFT analysis with the SMEFT analysis on an equal footing,
we turn off the finite temperature corrections from new physics in this analysis.
The red, blue, blue dashed, and blue dotted lines correspond to the results in 
our EFT, the EFT truncated up to mass dimension six, eight and ten, respectively.

We find that, if we compute $v_n/T_n$ by using the dimension six effective potential (\ref{eq:VnaHEFT-dim6}), we find that $v_n/T_n>1$ requires $r\gtrsim 0.35$ 
when $\Lambda = 1\,\mbox{TeV}$ and $\kappa_0=1$. 
This requirement corresponds to $f \lesssim 890\,\mbox{GeV}$,
which is consistent with the findings in the literatures \cite{
Grojean:2004xa,
Ham:2004zs,
Bodeker:2004ws,
Delaunay:2007wb,
Ellis:2019flb}.
It should be noted that, however, the dimension six approximation seems to fail because the prediction largely deviated from our EFT result which agrees with many concrete new physics model studies. 
We find $\simeq 44\%$ discrepancy when $r=0.35$, $\Lambda=1\,\mbox{TeV}$, and $\kappa_0=1$.
To reduce the discrepancy, 
we need to include the higher mass dimension operators to the EFT consisted of up to mass dimension six operators. 
For example, when $0.25\lesssim r\lesssim 0.5$, 
we can improve the EFT prediction by adding the higher mass dimension operators appropriately.
The importance of the higher dimensional operator for studying the strongly first-order EWPT has been also emphasized in Refs.~\cite{
Damgaard:2015con,
deVries:2017ncy,
Chala:2018ari,
Croon:2020cgk,
Postma:2020toi}.
However, when $r\gtrsim 0.5$, 
the improvement with finite number of the higher dimensional operators does not work. 
This is because the Taylor expansion (\ref{eq:log-expand}) becomes ill-defined due to $x_\Phi\gtrsim 1$ in this regime. 
Therefore, for $r\gtrsim 0.5$, we have to employ our EFT formalism, 
not the EFT formalism with the finite truncation of higher dimensional operators.

\section{Gravitational waves from the strongly first-order EWPT}
\label{sec:GWs}
We next estimate the GW spectra from the strongly first-order EWPT, which should be
one of the important signatures of our scenario as we mentioned in the introduction.
The spectrum of the GWs from the strongly first-order EWPT is characterized by two parameters, $\alpha$ and $\tilde{\beta}$. These parameters are defined by \cite{Grojean:2006bp}
\begin{align}
&\alpha
\,=\,
\frac{1}{\rho_{\rm{rad}}}
\biggl[
-\Delta V_{\rm{EFT}}
+T\frac{\partial \Delta V_{\rm{EFT}}}{\partial T}
\biggr]
\biggl|_{T=T_n}
\,,\\
&\tilde{\beta}
\,=\,
T\frac{d}{dT}\l(\frac{S_3}{T}\r)
\biggl|_{T=T_n}
\,,
\end{align}
where $\Delta V_{\rm{EFT}}=V_{\rm{EFT}}(\phi^B(T),T)-V_{\rm{EFT}}(0,T)$ with $\phi^B(T)$ being the bounce solution for the vacuum bubbles. 
$S_3$ is the free energy of the vacuum bubbles. 
$\rho_{\rm{rad}}$ denotes the radiation energy density, which is calculated as
\begin{align}
\rho_{\rm{rad}} (T)
\,=\,
\frac{\pi^2}{30}\,g_*(T)\,T^4\,,
\end{align}
where $g_*$ is the relativistic degrees of freedom. In our analysis, we simply neglect the temperature dependence of  $g_*$ and take $g_*=106.75+\kappa_0$.
The parameter $\alpha$ characterizes the latent heat, while the parameter $\tilde{\beta}$ corresponds to the inverse of the time duration of the phase transition (normalized by the Hubble scale). 
Typically, the stronger first-order phase transition predicts larger $\alpha$ and smaller $\tilde{\beta}$.

\begin{figure}[t]
	\centering
	\includegraphics[width=7cm,clip]{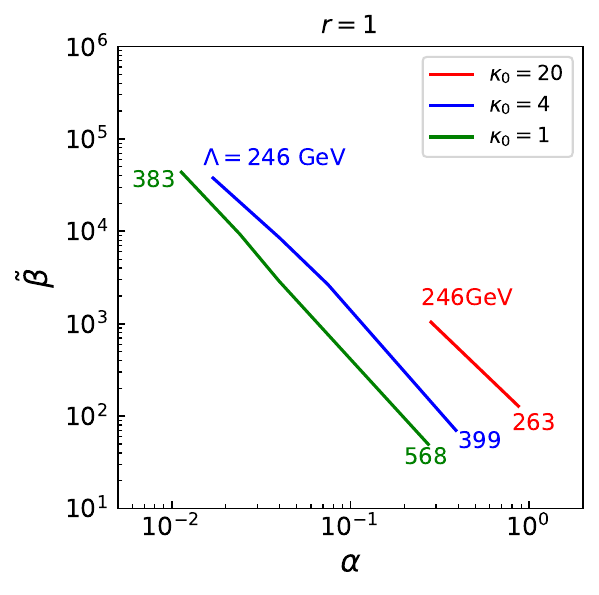}	
	\includegraphics[width=7cm,clip]{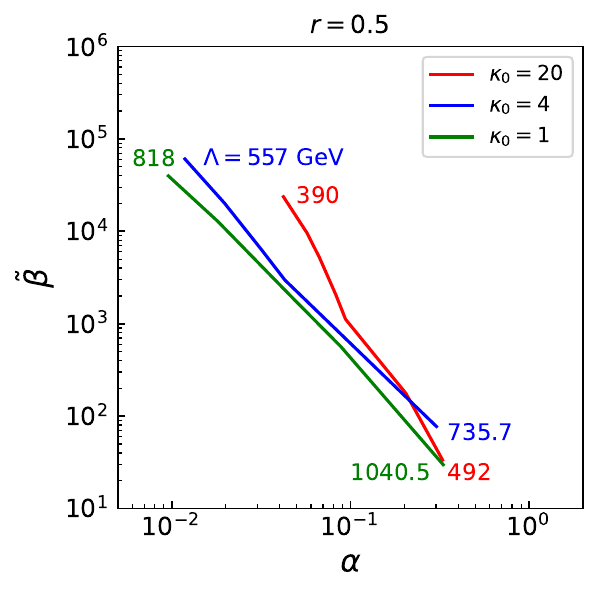}		
	\caption{ 
	    $\alpha$ and $\tilde{\beta}$ in our EFT.
	    We fix $r=1$ and $0.5$ in the left and right figures, respectively.
	    The green, blue and red lines correspond to $\kappa_0=1,4$, and $20$ cases. 
	    The numbers in these figures are the value of $\Lambda\,\mbox{[GeV]}$.
	    				}
	\label{fig:alpha-beta}
\end{figure}

The GW spectra ($\Omega_{\rm{GW}}$) produced by the first-order phase transition have three sources \cite{Caprini:2015zlo}; collisions of the vacuum bubbles $( \Omega_\varphi)$, compressional waves (sound waves) $(\Omega_{\rm{sw}})$ and magnetohydrodynamics turbulence $(\Omega_{\rm{turb}})$;
\begin{align}
h^2\Omega_{\rm{GW}}
\,\simeq\,
h^2\Omega_{\rm{\varphi}}
+
h^2\Omega_{\rm{sw}}
+
h^2\Omega_{\rm{turb}}
\,.
\end{align}
We estimate the GW spectra according to Ref.~\cite{Caprini:2015zlo}.
The GW spectra can be estimated once we determine $\alpha$, $\tilde{\beta}$, and the bubble wall velocity ($v_b$). 
We take $v_b=0.95$ in our analysis.

\begin{figure}[t]
	\centering
	\includegraphics[width=7cm,clip]{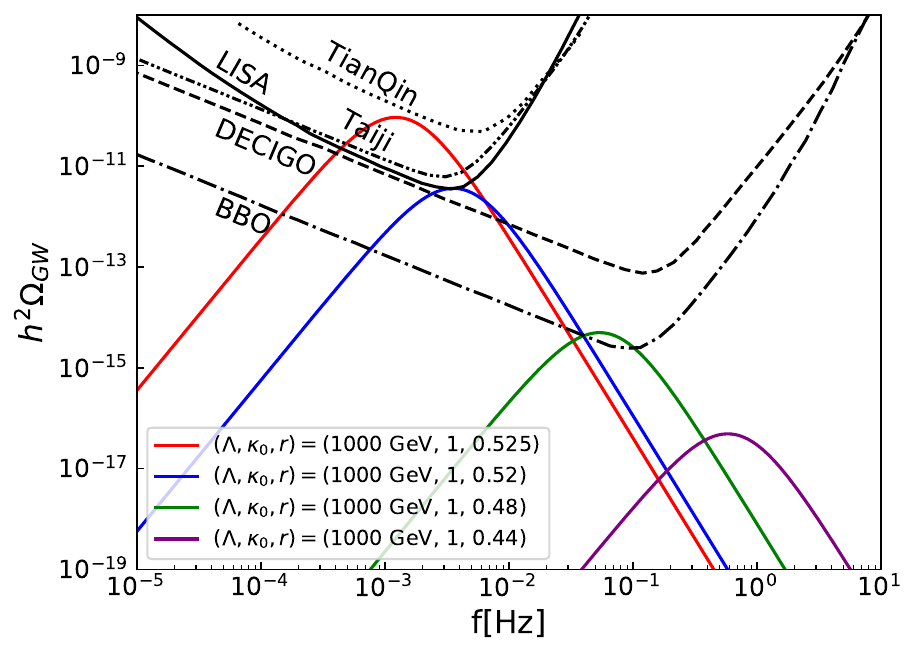}
	~~
	\includegraphics[width=7cm,clip]{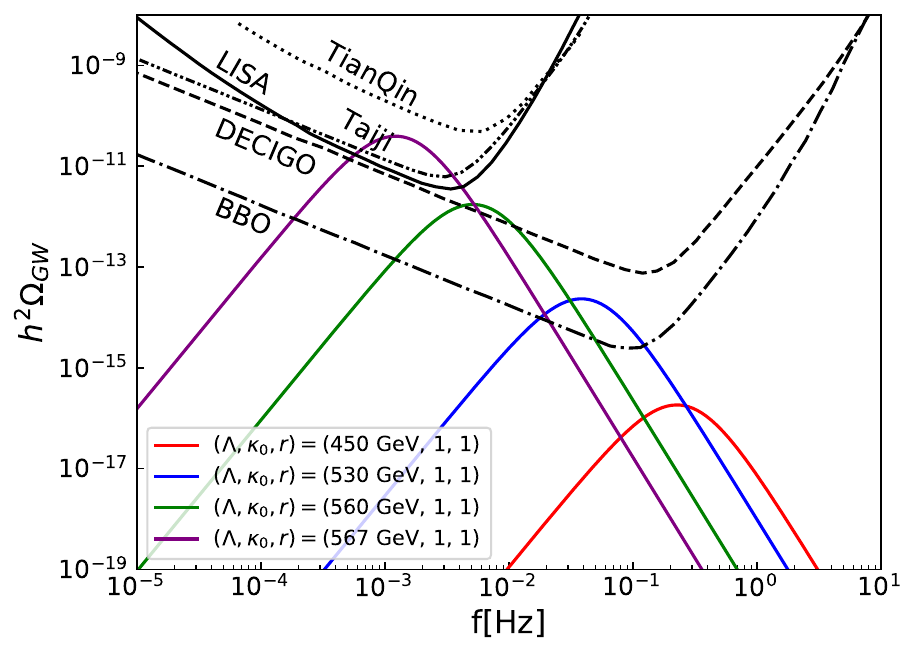}		
	~~
	\includegraphics[width=7cm,clip]{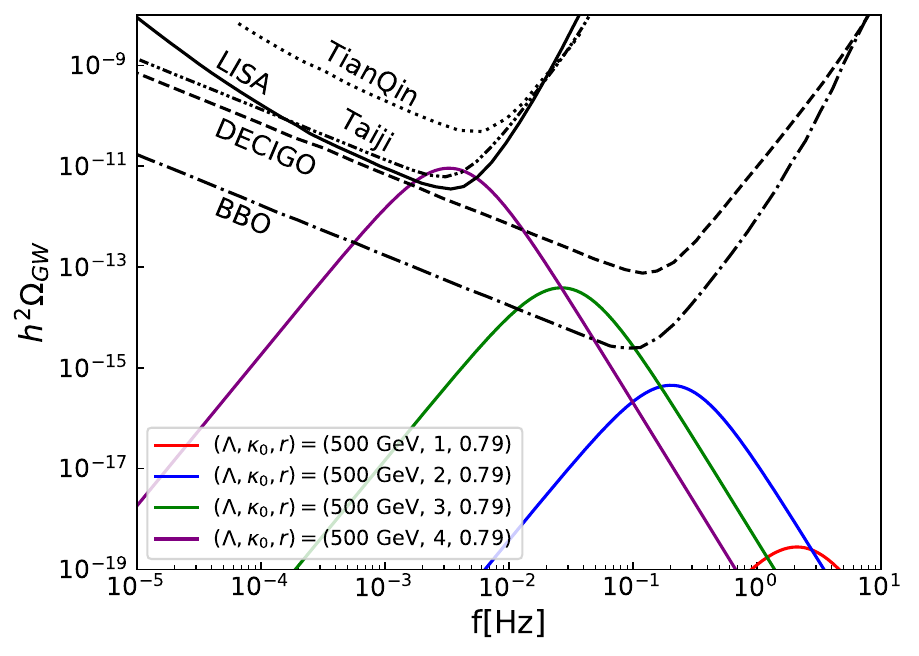}
	\caption{ 
	    The GW spectra for some benchmark values.
	   				}
	\label{fig:GWspectra-naHEFT}
\end{figure}

In figure \ref{fig:alpha-beta}, 
we show the predicted values of $\alpha$ and $\tilde{\beta}$ with varying $\Lambda$ and $\kappa_0$. The green, blue, and red lines correspond to the case for $\kappa_0=1, 4,$ and $20$, respectively. The numbers in these figures are the values of $\Lambda\,\mbox{[GeV]}$.
We set $r=1$ and $0.5$ in the left and right figures, respectively.
The maximum value of $\Lambda$ is determined by the condition of the completion of the phase transition. The minimum value of $\Lambda$ is, on the other hand, fixed by the condition of the realization of the strongly first-order EWPT or $\Lambda\geq v$. 
We find that the larger $\Lambda$ predicts the larger $\alpha$ and smaller $\tilde{\beta}$. 
This is because, in the case with sizable $r$ (corresponding to the non-decoupling case), the large $\Lambda$ induces the significant non-decoupling effect which results in large $v_n/T_n$.

We show the GW spectra for some benchmark values in figure \ref{fig:GWspectra-naHEFT}. 
The colored lines are the predicted GW spectra, while the black lines are 
the sensitivity curves of the LISA \cite{LISA:2017pwj}, DECIGO \cite{Seto:2001qf},  TianQin \cite{TianQin:2015yph}, Taiji \cite{Ruan:2018tsw} and BBO \cite{Corbin:2005ny}{\footnote{
Performing the detailed analysis of the sensitivity, 
one can find the factor improved effective sensitivity comparing than ones shown in figure \ref{fig:GWspectra-naHEFT} \cite{Hashino:2018wee}.}}. 
In the top-left figure, we show the $r$-dependence in the GW spectra.
Here we take $\Lambda=1\,\mbox{TeV}$ and $\kappa_0=1$. 
The red, blue, green, and purple lines correspond 
to the results for $r=0.525$, $0.52$, $0.48$, and $0.44$, respectively. 
We find that the larger $r$ makes the amplitude and the peak frequency larger and lower. 
This is because the large $r$ makes the first-order phase transition stronger, and results in the large $\alpha$ and small $\tilde{\beta}$.  
We also show the $\Lambda$- and $\kappa_0$-dependences in the top-left and the bottom figures, respectively. 
We find that the larger $\Lambda$ and $\kappa_0$ also make the height of the spectra and the peak frequency higher and lower. 
This is due to the same reason as the large $r$ case.

\section{Discussion}
\label{sec:result}
\begin{figure}[t]
	\centering
	\includegraphics[width=7cm,clip]{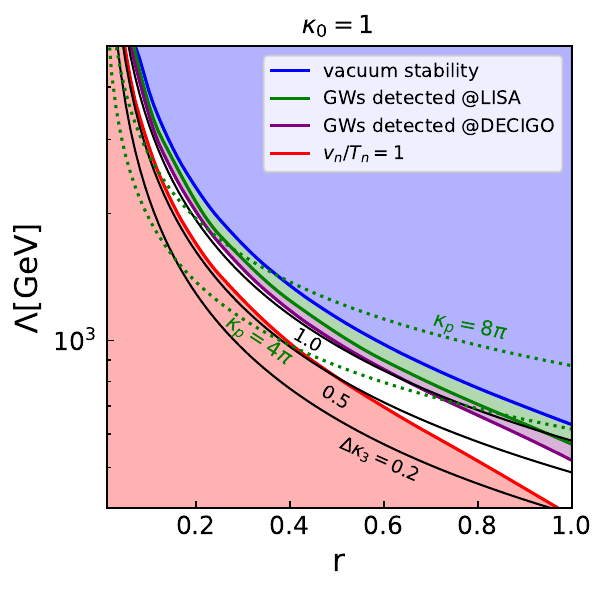}
	~~
	\includegraphics[width=7cm,clip]{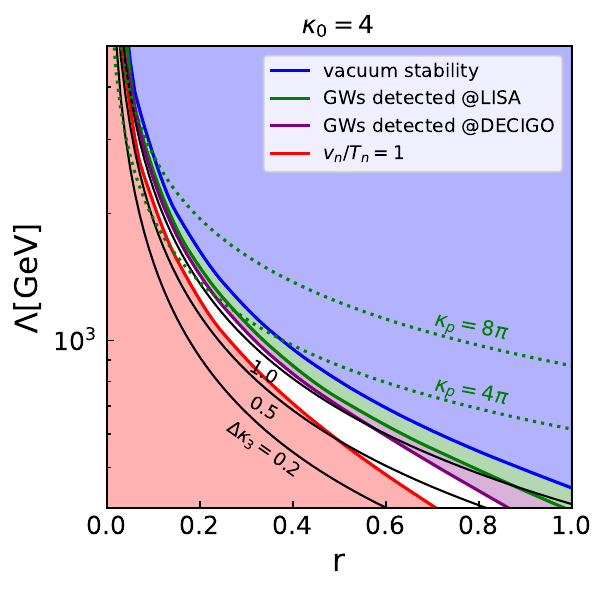}		
	\caption{ 
	   Parameter space in which the GW will be detected at LISA (green) and DECIGO (purple).
	   The color notation of the red and blue regions are same with figure \ref{fig:SFOEWPT}.
	   We take $\kappa_0=1$ and $4$ in the left and right figure, respectively.
	   The black lines are contours of $\Delta\kappa_3=\kappa_3-1=0.2,0.5$ and $1$.
	   	   				}
	\label{fig:result-Lambda-r}
\end{figure}
We finally discuss the predictions of our scenario by combining all observations we obtained so far. 
The final result is summarized in figure \ref{fig:result-Lambda-r}, in which we add the information of the sensitivity of the future GW observations and the deviation of the $hhh$ coupling to figure \ref{fig:SFOEWPT}.
The green and purple regions respectively correspond to the parameter region where the predicted GW spectrum is above the sensitivity curves of LISA and DECIGO shown in figure \ref{fig:GWspectra-naHEFT}. 
We also show the predicted $hhh$ coupling deviation factors $(\Delta\kappa_3=\kappa_3-1)$ by the black lines, which correspond to $\Delta \kappa_3=0.2,0.5$ and $1$.
We estimate $\Delta \kappa_3$ by using Eq.~(\ref{eq:kappa3}).
In addition to the GW observations, 
the $hhh$ coupling measurements at future colliders are also important for investigating
 the parameter region for the strongly first-order EWPT. 
As we see, the parameter region for strongly first-order EWPT predicts $\mathcal{O}(10\,\%) \sim \mathcal{O}(100\,\%)$ deviation in $hhh$ coupling, and such large deviation is expected to be tested by the future collider experiments. For example, the HL-LHC could reach an accuracy of 50\% \cite{Cepeda:2019klc}, and the lepton colliders such as the ILC with $\sqrt{s}\simeq 1\,\mbox{TeV}$ is expected to achieve a precision of some tens of percent \cite{Fujii:2017vwa,Roloff:2019crr}. 

In summary, we find that 
the scenarios in which the strongly first-order EWPT is realized by the non-decoupling quantum effects can be broadly tested by the future GW interferometers and the precise measurement of the $hhh$ coupling at the future colliders.
This findings has been pointed out in many concrete model studies so far.
See Refs.~\cite{Kakizaki:2015wua, Hashino:2016rvx} for example.  
Our EFT analysis represents the essential consequences found in these model-dependent studies.

\section{Summary and conclusions}
\label{sec:conclusion}
We provided an effective field theoretical description of the strongly first-order EWPT.
We employed an extended Higgs effective field theory in which the Higgs potential is parameterized by a Coleman-Weinberg like form.
Our EFT systematically describes a class of new physics models in which i) the mixing between new scalar particles and the SM Higgs boson is much suppressed and ii) the sizable non-decoupling quantum effect appears. 
We formulated the finite temperature corrections in the EFT, and studied the EWPT by using the EFT framework.

We found that the non-decoupling effect plays an essential role in realizing the strongly first-order EWPT. This observation is consistent with the findings in many concrete models studies. 
The importance of the non-decoupling effect means that 
the conventional EFT description with the finite number truncation of the higher-dimensional operators is not appropriate for study on the strongly first-order EWPT. 
Comparing our EFT results with the ones estimated by the effective potential with finite number truncation of higher dimensional operators,
we numerically showed that the approximation of the finite number truncation does not work in the typical parameter space of the strongly first-order EWPT.

Applying our EFT formalism, 
we estimated the spectrum of the stochastic GWs produced at the strongly first-order EWPT.
We found that the non-decoupling effect results in the GW spectrum which is expected to be detected at the future GW interferometers like LISA and DECIGO.
We also emphasized that the deviation of the triple Higgs coupling is also an important signature of our scenario.
$\mathcal{O}(10\,\%) \sim \mathcal{O}(100\,\%)$ deviation of the triple Higgs coupling is predicted in the parameter space where the strongly first-order EWPT is realized.
This observation means that the future $hhh$ couplings measurement and the GW observations are important probes for the scenarios with the strongly first-order EWPT. 
Our findings can be generally applied to the class of new physics models in which the strongly first-order EWPT is realized by non-decoupling quantum effects.

We finally emphasize the importance of the study on non-decoupling physics. The non-decoupling effects should be essential origin of the strongly first-order EWPT, and provide us rich phenomenology. 
It is therefore important to study these non-decoupling new physics in a complemental way. 
Our EFT description should be one of the simplest and most systematic descriptions for this approach. 
The further phenomenological analysis based on our EFT is performed elsewhere.

\section*{Note added}
While this work has been finalized, 
a related analysis has been reported in Ref.~\cite{Banta:2022rwg},
in which the impact of the non-decoupling effects to the electroweak phase transition is studied by considering several concrete extended scalar models.

\section*{Acknowledgments}
This work of S.K. was supported, in part, by the Grant-in-Aid on Innovative Areas, the Ministry of Education, Culture, Sports, Science and Technology, No. 16H06492 and by the JSPS KAKENHI Grant No. 20H00160. 
The work of R.N. was supported by JSPS KAKENHI Grant No. JP19K14701 and JP21J01070. 
The work of M.T. was supported in part by JSPS KAKENHI Grant No. JP21J10645.
\bibliography{draft} 
\bibliographystyle{JHEP}
\end{document}